%Summary of string functions GRR
% Initiated Rehovot 18/9/14
\input phyzzx
\def\e{\adveq\eqno{\rm (\chapterlabel\the\equanumber)}}

\def\adveq{\global\advance\equanumber by 1}
\def\myeq{{\rm \chapterlabel.\the\equanumber}}
\def\rarrow{\rightarrow}

\def\twoline#1#2{\displaylines{\qquad#1\hfill(\adveq\myeq)\cr\hfill#2
\qquad\cr}}

\def\semidirect{\mathrel{\raise0.04cm\hbox{${\scriptscriptstyle |\!}$
\hskip-0.175cm}\times}}

%\define\semidirect{\propto}
\def\mod{\mathop{\rm mod}\nolimits}

\def\ref#1{$^{[#1]}$}

\def\r#1{$[\rm#1]$}

\def\Tr{\mathop{\rm Tr}\limits}
\def\e{\adveq\eqno{\rm (\chapterlabel\the\equanumber)}}

\def\adveq{\global\advance\equanumber by 1}
\def\myeq{{\rm \chapterlabel\the\equanumber}}
\def\rarrow{\rightarrow}

\def\twoline#1#2{\displaylines{\qquad#1\hfill(\adveq\myeq)\cr\hfill#2
\qquad\cr}}

\def\semidirect{\mathrel{\raise0.04cm\hbox{${\scriptscriptstyle |\!}$
\hskip-0.175cm}\times}}

%\define\semidirect{\propto}
\def\mod{\mathop{\rm mod}\nolimits}

\def\ref#1{$^{[#1]}$}

\def\r#1{$[\rm#1]$}

\def\Tr{\mathop{\rm Tr}\limits}

\def\half{{1\over2}}

\overfullrule=0pt
\date{October, 2014}
\date{October, 2014}
\titlepage
\title{On The Characters of Parafermionic Field Theories}
\author{Doron Gepner}
\vskip20pt
\line{\it\hfill  Department of Particle Physics, Weizmann Institute, Rehovot, Israel\hfill} 

\abstract
We study cosets of the type  $H_l/U(1)^r$, where $H$ is any Lie algebra at level $l$
and rank $r$. 
These theories are parafermionic and their characters are related to the string functions,
which are generating functions for the multiplicities of weights in the affine representations.
An identity for the characters is described, which apply to all the algebras
and all the levels. 
The expression is of the Rogers Ramanujan type.
We verify this conjecture, for many algebras and levels, using
Freudenthal Kac formula, which calculates the multiplicities in the affine 
representations, recursively, up to some grade.
Our conjecture  encapsulates all the known results about these string functions,
along with giving a vast wealth of new ones. 
\endpage 
Some years ago the author have suggested parafermionic conformal field theories related 
to simple Lie algebras
\REF\New{D. Gepner, Nucl.Phys. B 290 (1987) 10.}\r\New.
The characters of these conformal field theories (CFT) were
shown to be expressible in terms of the string functions of the corresponding affine Lie
algebra (up to factors of Dedekind's eta function).
Our purpose here is to describe closed expressions for many of these characters,
which are of the Rogers Ramanujan type. 

Several examples of  characters in CFT were studied, and were  shown to be
expressible as Rogers Ramanujan type sums 
\REF\LP{J. Lepowski and M. Primc, Contemporary Mathematics vol. 46 (AMS, Providence, 1985).}
\REF\Das{V. S. Dasmahapatra, T. R. Klassen, B. M. McCoy and E. Melzer,
J. Mod. Phys. B7 (1993) 3617}
\REF\Kedem{R. Kedem, T. R. Klassen, B. M. McCoy and E. Melzer, Phys. Lett. B 307
(1993) 68.}
\REF\Baver{E. Baver and D. Gepner, Phys. Lett. B 372 (1996) 231.}
 \REF\Kuniba{A. Kuniba, T. Nakanishi and J. Suzuki, Mod. Phys. Lett A8 (1993)1649.}
 \REF\BG{A. Belavin
and D. Gepner, Lett. Math. Phys. 103 (2013) 1399.}
\REF\GG{A. Genish and D. Gepner, Level two string functions and Rogers Ramanujan type identities, arXiv:1405.1387, Nucl. Phys. B, in press.}
\r{\LP,\Das,\Kedem,\Baver,\Kuniba,\BG,\GG}. The origin of these identities is somewhat perplexing,
though, they were shown to be connected with local state probabilities in solvable lattice
models 
\REF\Baxter{R. J. Baxter, Exactly solved models in statistical mechanics (Dover books
on physics).}
\REF\Melzer{E. Melzer, Int. J. Mod. Phys. A 9 (1994) 1115.}
\REF\Lat{D. Gepner, Phys. Lett. B 348 (1995) 377.}
\r{\Baxter,\Melzer,\Lat}, 
and with thermodynamic Bethe ansatz equations (see, e.g.,  \r\Kuniba\ and refs. therein).

The characters of the parafermionic field theories are 
$\chi^\Lambda_\lambda(\tau)$, where $\Lambda$ is an integrable dominant weight of some 
simple Lie algebra $H$ at level $l$, and
$\lambda$ is the weight. We describe here a Rogers Ramanujan type expression when
the weight $\Lambda=f\Lambda_g$ where  $\Lambda_g$ has mark one, and $f$ is any integer,
or $\Lambda_g$ has dual mark one and $f=1$.  Our  conjecture holds for any 
allowed weight $\lambda$, and for all algebras $H$ at any level $l$. 

We verify our conjecture using Freudenthal Kac formula, which gives the exact characters
up to a certain grade (or the dimension of the fields). We use for this purpose a computer 
program that we wrote.

This provides important physical information about the parafermionic theories, along
with valuable mathematical insight concerning the string functions, which are central in
the mathematics of infinite dimensional Lie algebras.

Parafermionic conformal field theories associated with Lie algebras were described in ref.
\r\New, along with their characters
and partition functions. These theories can be thought of as the cosets of the type,
$${H_l\over U(1)^r},\e$$
where $H$ is  any Lie algebra, of the types $A-G$,  $l$ is the  level and $r$ is the rank.

The fields in the theory are labeled by a pair of weights $(\Lambda,\lambda)$, where $\Lambda$ 
is a integrable  highest weight of $H$ at level $l$ and $\lambda$ is an element of the weight
lattice of $H$. We have the selection rule,
$$\Lambda-\lambda\in M,\e$$
where $M$ is the root lattice of $G$. This lattice is generated by the simple roots, which we
denote by $\alpha_a$, $a=1,2,\ldots,r$.
The characters of the theory are denoted  by $\chi^\Lambda_\lambda(\tau)$ and they can
be expressed in terms of the string functions of the corresponding untwisted affine Lie
algebra, as follows,
$$\chi^\Lambda_\lambda(\tau)=\eta(\tau)^r c^\Lambda_\lambda(\tau),\e$$
where $\eta(\tau)$ is the Dedekind's eta function.  The string functions
are defined as \REF\KP{V. Kac and D. Petersen, Adv. Math. 53 (1984) 125.}
\r{\KP,\LP},
$$\twoline{c^\Lambda_\lambda(\tau)=e^{-\pi i\lambda^2\tau/l}\Tr_{{\cal H}^\Lambda_\lambda} e^{2\pi i \tau (L_0-c/24)}=}{
e^{2\pi i\tau[\half (\Lambda+2\rho)\Lambda/(l+g)-\lambda^2/(2l)-c/24]}\sum_{n=0}^\infty
 p_n e^{2\pi i n\tau},}$$
where $L_0$ is the dimension of the fields,  $c=lD/(l+g)$ is the central charge, where $D$ is
the dimension of the algebra, and  ${\cal H}^\Lambda_\lambda$ is the 
representations of the affine algebra with dominant highest weight $\Lambda$ and 
weight $\lambda$. Here $\rho$ is half the sum of positive roots and  $g$ is the dual Coxeter number,
of the algebra $H$.
The integer $p_n$ is the multiplicity of the fields with the number operator $N=n$ (or the grade),
in the representation of the affine algebra  $\hat H$, which have the highest weight $\Lambda$ at level $l$, and  have a weight $\lambda$. Thus, the string functions are the generating functions
for the multiplicities of a 'string' of weights. 

The characters of the CFT (and the stirng functions) obey the following relations.
$$ \chi^\Lambda_\lambda(\tau)=\chi_{w(\lambda)}^\Lambda(\tau)=\chi^\Lambda_{\lambda+\mu}(\tau),\e$$
where $\omega$ is any element of the Weyl group of the finite algebra $H$, denoted by $W$, and
$\mu\in l M_L$, where $M_L$ stands for the long  root lattice of $H$, generated by the elements
${2\alpha_a/ \alpha_a^2}$, for $a=1,2,\ldots, r$. 

Before getting to our conjecture for the characters, we need to introduce some notation.
Let $t_a=2/\alpha_a^2$. Denote also $l_a=t_a l$, where $l$ is the level. We define 
$$G=\{ (a,m)  | \ 1\leq a \leq r,\quad 1\leq m\leq l_a-1,\quad  a,m\in Z\},\e$$
following ref.
\r\Kuniba.

We define 
$$K^{mk}_{ab}=\left({\rm min}(t_b m, t_a k)-{m k\over l}\right) \alpha_a\cdot \alpha_b.\e$$
Let $C_N$ be the Cartan matrix of $SU(N)=A_{N-1}$,
$$(C_N)_{r,s}=2\delta_{r,s}-\delta_{r,s+1}-\delta_{r+1,s},\e$$
where $r,s=1,2,\ldots, N-1$. We denote the inverse matrix by
$$B_N=(C_N)^{-1}.\e$$

Our Gerneralized Rogers Ramanujan expression (GRR)  involves a summation over the vector of non negative integers,
$${\bf n}=\left( n^{(a)}_m\right )_{(a,m)\in G}.\e$$
We also define the element of the root lattice,
$$\lambda({\bf n})=\sum_{(a,m)\in G} m n^{(a)}_m \alpha_a.\e$$
and the Pochhammer symbol,
$$(q)_{\bf n}=\prod_{(a,m)\in G} (q)_{n_m^{(a)}},\qquad (q)_k=\prod_{j=1}^k (1-q^j).\e$$
We also define
$${\cal K}({\bf n})={1\over2} \sum_{(a,m)\in G \atop (b,k)\in G} K^{mk}_{ab} n_m^{(a)} n_k^{(b)}.\e$$

A conjecture for the characters of the type $\chi^0_\lambda(\tau)$ was described by Kuniba et al.
\r\Kuniba. Their conjecture is
$$\chi^0_\lambda(\tau)=q^{-c_{\rm pf}/24} \sum_{\lambda({\bf n})=\lambda \mod l M_L}
{q^{{\cal K}({\bf n})}\over (q)_{\bf n}},\e$$
where $c_{\rm pf}$ is the central charge of the parafermions, $c_{\rm pf}=c-r$. 
There is a summation over the integers ${\bf n}$ from zero to infinity,
and
$$q=e^{2\pi i \tau}.\e$$
This conjecture holds for all the algebras $H$ and all the levels $l$ and any weight $\lambda$
on the root lattice $M$.
Our aim  here is to verify this conjecture, as well as establishing  expressions for the  characters $\chi^\Lambda_\lambda(\tau)$, such that $\Lambda\neq 0$.

Our conjectured identities holds for all the characters $\chi^\Lambda_\lambda(\tau)$, where the  weight
 $\Lambda$ is given by 
$$\Lambda=f \Lambda_g,\e$$ 
where  $f$ is a non-negative integer and $\Lambda_g$ is the $g$th  fundamental weight, and
for any  weight $\lambda$, which obeys the admissibility condition $\Lambda-\lambda\in M$.
We also assume that $\Lambda_g$ has mark one, i.e., $\Lambda_g\cdot\theta=1$, where
$\theta$ is the highest root and $\alpha_g^2=2$. Our conjecture also holds for fundamental weights which have 
dual mark one, $\Lambda_g\cdot \theta=1$ assuming $\alpha_g$ is a short root,
provided we take, $f=1$. 

For such weights $\Lambda=f\Lambda_g$, as above, we define
$${\cal L}_{f,g}({\bf n})=\sum_{m=1}^{l_g-1} (B_{l_g})_{l_g- f,m } n^{(g)}_m,\e$$
when $1\leq f\leq l_g-1$ and we otherwise assume ${\cal L}_{f,g}({\bf n})=0$. The matrix $B_N$ was
defined in eq. (9).

Our main result, which is a conjecture for $\chi^\Lambda_\lambda(\tau)$, is then
$$\chi^\Lambda_\lambda(\tau)=q^{-\Delta^\Lambda_\lambda} \sum_{\lambda({\bf n})=\lambda-\Lambda \mod l M_L} {q^{-{\cal L}_{f,g}({\bf n})+{\cal K}({\bf n})}\over (q)_{\bf n}},\e$$
where there is a summation of ${\bf n}$ over all the non negative integers,
${\bf n}\in (Z_{\geq0})^{|G|}$,
and  $\Delta^\Lambda_\lambda$ is some dimension which we do not specify. The GRR formula
eq. (18) holds for all the algebras, $H$, at all the levels.

This identity is, by definition, invariant under the symmetry, eq. (5), where we take $\lambda\rarrow
\lambda+\mu$, where $\mu\in l M_L$, which is a good consistency check. It is also identical  when $\Lambda=0$
to the conjecture of Kuniba et al., eq. (14). In the case of $H=SU(2)=A_1$, a GRR expression for
the characters was obtained by Lepowski et al. \r\LP, and it agrees  exactly with our conjecture. 
Also, for the level two simply laced algebras a GRR expression was described in 
refs. \r{\BG,\GG},
and our formula here specialises precisely
to these level two results. Thus our formula, eq. (18), encapsulates
all the known GRR expressions for the characters of the parafermionic field theories, along with providing 
a wealth of new ones.

The characters are invariant under the field identifications \r\New,
$$\chi^\Lambda_\lambda(\tau)=\chi^{\sigma(\Lambda)}_{\sigma(\lambda)}(\tau),\e$$
where $\sigma$ is any automorphism of the affine Dynkin diagrams, and we assumed that
$\lambda$ is also an integrable weight at level $l$. When $\Lambda=f\Lambda_g$,
where $\Lambda_g\cdot \theta=1$, and $\alpha_g^2=2$, this is equivalent to taking $f$ to $l-f$, and eq. (19)  is indeed
a symmetry of the GRR expression eq. (18),  (when we transform also $\lambda$ appropriately), providing a consistency check for this identity.  

The conjecture eq. (14) was shown in ref. \r\Kuniba\ to give the correct central charges by 
evaluating the
$q\rarrow 1$ limit, making connection with thermodynamic Bethe ansatz equations .
Our conjecture eq. (18) thus also gives the correct central charges since
the linear terms that we added do not change the calculation of the central charges,  as they can be neglected when the elements of  ${\bf n}$ are all large.

The symmetry of the characters under the Weyl group, eq. (5), $\chi^\Lambda_\lambda(\tau)=
\chi^\Lambda_{w(\lambda)}$, where $w$ is any element of the Weyl group, is not
explicit in the GRR, eq. (18), but actually gives an expression with a different summation condition,
for each $w\in W$. These GRR sums are, of course, equal, by eq. (5), and gives a wealth of nontrivial identities, which we term  generalised Slater identities, as the simplest case,
for the Ising model, was described by Slater \REF\Slater{L. J. Slater, Proc. Lond.  Math. Soc. 54 (1953) 147.}\r\Slater. In special cases, such identities were noted already in refs. \r{\BG,\GG}.

We get now to the problem of verifying our general GRR identity eq. (18). The characters of the 
parafermionic theories are closely related to the string functions, eq. (3). These string functions
are, actually, multiplicities of weights in the corresponding integrable affine highest weight representation,
eq. (4). We denote this representation by $L(\bar\Lambda)$ where $\bar\Lambda$ is the
affine highest weight, whose finite weight is $\Lambda$ and its  level is $l$.
For explanation of this notions see the book by Kac \REF\KBook{V. G.  Kac, Infinite dimensional Lie algebras, Cambridge Univ. press, 1990.}\r\KBook, or for a review, the appendix of ref.
\REF\GW{D. Gepner and E. Witten, Nucl. Phys. B 278 (1986) 493.}\r\GW. 
Thus, the problem of calculating the characters reduces to the problem of calculating the multiplicities of weights
in the appropriate representation of the affine algebra $\hat H$. Luckily, a great deal of results
are known about these. We find, of particular suitability, the generalisation by Kac 
\r\KBook, p. 211, of Freudenthal's formula \REF\Humphrey{
J. E. Humphreys, Introduction to Lie algebras and representation theory. Graduate texts in
mathematics, Springer (1997).}
\r\Humphrey,
which is valid for any Kac Moody algebra, and  is 
$$\left( |\bar \Lambda+\bar \rho|^2-|\bar \lambda+\bar\rho|^2\right) {\rm dim}V_{\bar\lambda}=
2\sum_{\alpha\in \Delta_+}\sum_{j\geq 1} ({\rm mult\,}\alpha)( \bar\lambda+j\alpha|\alpha){\rm
dim} V_{\bar \lambda+j\alpha}.\e$$ 
We specialise here to the untwisted affine algebra $\hat H$. Here $\bar\Lambda$ and 
$\bar \lambda$ are the affine weights, $\bar\rho$ is the affine generalisation of half the  sum
of positive roots,  $\Delta_+$ is the set of positive affine roots and ${\rm mult\, \alpha}$ is the 
multiplicity of the root $\alpha$. The scalar product $(a | b)$
is the affine scalar product and we denote by ${\rm dim}V_{\bar\lambda}$ the dimension
of the vector space with the weight $\bar\lambda$ (or, multiplicity). 

The Freudenthal Kac formula, eq. (20), is an effective tool for calculating the multiplicities in
the affine representation, as it can be used recursively, for each representation, starting from the highest weight,
grade by grade. For a discussion and examples for simple Lie algebras see Humphrey's book \r\Humphrey.
For affine algebras, since the representation is infinite,  we simply have to stop at some grade
(or, dimension). This sort of computation, can be done by a computer program. We implemented 
this algorithm in 
the fortran program ALGEBRA (written by A. Abouelsaood and D. Gepner),
which calculates the multiplicities either for finite or untwisted affine algebras.  

Let us give some examples. We find it convenient to rearrange the elements of $K$ in a matrix, 
$$M_{r,s}=K^{G[r,2],G[s,2]}_{G[r,1],G[s,1]},\e$$
where $G$ and $K$ were defined in eqs. (6,7), and  $r,s=1,2,\ldots, |G|$ and we denote by $G[r,k]$ the $k$th entry of the $r$th element of $G$. 

Consider the algebra $B_2$ at level two. The simple roots of the algebra are 
$\alpha_1=\epsilon_1-\epsilon_2$ and $\alpha_2=\epsilon_2$, where $\epsilon_i$ are orthogonal
unit vectors. The fundamental weights are $\Lambda_1=\epsilon_1$ and $\Lambda_2=(\epsilon_1+\epsilon_2)/2$. Here $l_1=2$,  $l_2=4$, and 
$$M=\half\pmatrix{2 & -1 & -2 &-1\cr
                              -1 &3&2 &1\cr
                              -2 &2&4&2\cr
                              -1 & 1 & 2 & 3\cr}.\e$$ 
 We define the vector $\bf n$ as, eq. (10),
 $${\bf n}=(n_1,n_2,n_3,n_4)=(n_1^{(1)},n_1^{(2)},n_ 2^{(2)},n_3^{(2)}).\e$$
 We consider the fundamenal weight of the short root, $\Lambda_2=(\epsilon_1+\epsilon_2)/2$.
 For $\lambda$ we consider two cases.  First, for $\mu=0$ we take $\lambda=\Lambda_2$,
 and for $\mu=1$ we take $\lambda=\Lambda_2+\alpha_2=(\epsilon_1+3\epsilon_2)/2$.  
 
 Then, our general GRR identity, eq. (18) becomes,
 $$z_\mu=q^{\Delta_\mu} \chi^{\Lambda_2}_{\Lambda_2+\mu\alpha_2}=
 \sum_{n_1=0\mod 2\atop n_2+2 n_3+3 n4=\mu\mod 4}      
 {q^{{\bf n} M {\bf n}^t/2-(n_2+2 n_3+3 n_4)/4}\over (q)_{\bf n}},\e$$
 where there is a sum over all the $n_i$ from zero to infinity,
 and  $\Delta_\mu$ is set to make the leading term $q^{\mu/2}$.
 
 We can  evaluate the sum, eq. (24), by a Mathematica program and we find
 $$z_0= 1 + 3 q + 9 q^2 + 22 q^3 + 46 q^4 + 93 q^5 + 176 q^6 + 319 q^7 + 
 562 q^8 + 960 q^9 + O(q^{10}),\e$$
 and
$$q^{-\half} z_1=2+6 q+14 q^2+32 q^3+66 q^4+128 q^5+238 q^6+426 q^7+736 q^8+1242 q^9+O(q^{10}).\e$$

From the program ALGEBRA we can calculate  the relevant string functions
which are $c^{\Lambda_2}_{\Lambda_2+\mu\alpha_2}(\tau)$, by reading off the dimensions in the
representation $L(\bar\Lambda_2)$ at level two. We find, up to grade $8$,
$$ 
q^{d_0}c^{\Lambda_2}_{\Lambda_2}(\tau) = 1 + 5 q + 20 q^2 + 65 q^3 + 185 q^4 + 481 q^5 + 1165 q^6 + 
  2665  q^7 + 5822 q^8+O(q^9).\e$$
  Multiplying it by $\eta(\tau)^2$ according to eq. 
  (3), we find the character up to order $8$, and these agree exactly with $z_0$, calculated from the
  GRR, eq. (25). 
 For the other string function, we find from ALGEBRA,
 $$q^{d_1} c^{\Lambda_2}_{\Lambda_2+\alpha_2}(\tau)=
 2 + 10 q + 36 q^2 + 110 q^3 + 300 q^4 + 752 q^5 + 1770 q^6 + 3956 q^7+O(q^8),\e$$
 which, again, when multiplied by $\eta(\tau)^2$ agrees exactly with $z_1$.
 The quantities $d_\mu$ are some dimensions, which follow from eq. (4).
 Many more string functions can be compared with the GRR and indeed, they all agree, verifying
 our general conjecture eq. (18) for the algebra $B_2$ at level two.
 
 Let us give another example which is $SU(3)=A_2$ at level $3$. The simple roots
 of $A_2$ are $\alpha_1=\epsilon_1-\epsilon_2$ and $\alpha_2=\epsilon_2-\epsilon_3$.
 The fundamental weights are, $\Lambda_1=(2\alpha_1+\alpha_2)/3$ and 
 $\Lambda_2=(\alpha_1+2\alpha_2)/3$. Both weights have mark one. Here,
 $t_1=t_2=1$ and so $l_1=l_2=l=3$. The Cartan matrix is 
 $$C_{3}=\pmatrix{2&-1\cr -1& 2\cr},\e$$
 and
 $$G=\{(1,1),(1,2),(2,1),(2,2)\}.\e$$
 Here,
 $$\twoline{K_{a,b}^{m,k}= ({\rm min}(m,k)-m k/3) \alpha_a\cdot\alpha_b=}{
 (B_3)_{mk} (C_3)_{ab}=
 {1\over3} \pmatrix{2 &1\cr 1& 2\cr}_{m,k} \pmatrix{2 &-1\cr -1&2\cr}_{ab}.}$$
 Also,
 $${\bf n}=(n^{(1)}_1,n^{(1)}_2,n^{(2)}_1,n^{(2)}_2).\e$$
 From eq. (13),
 $${\cal K}({\bf n})={1\over2} \sum_{a,b,m,k=1}^2 K_{a,b}^{m,k} n^{(a)}_m n^{(b)}_k.\e$$
 
 We consider the highest weight $\Lambda_1=(2\alpha_1+\alpha_2)/3$ and
 the weights $\lambda=\Lambda_1+\mu\alpha_1$, where $\mu=0,1$.
 Our general  GRR, eq. (18), then becomes
 $$z_\mu=q^{d_\mu}\chi^{\Lambda_1}_{\Lambda_1+\mu\alpha_1}=\sum_{n^{(1)}_1+2 n^{(1)}_2=\mu \mod 3
 \atop
 n^{(2)}_1+2 n^{(2)}_2=0\mod 3}
 {q^{{\cal K}({\bf n})-(n^{(1)}_1+2 n^{(1)}_2)/3}\over (q)_{\bf n}},\e$$
 where $d_\mu$ is a dimension used to make the leading term $q^{\mu/3}$.
 
 Using a Mathematica program, we evaluate $z_\mu$, eq.  (34), and we find
 $$z_0=1 + 2 q + 7 q^2 + 16 q^3 + 36 q^4 + 70 q^5 + 135 q^6 + 243 q^7 + 
 431 q^8 + 731 q^9 + O(q^{10}).\e$$
 and
 $$q^{-{1\over 3}} z_1=1 + 4 q + 9 q^2 + 22 q^3 + 44 q^4 + 89 q^5 + 163 q^6 + 297 q^7 + 
 513 q^8 + 874 q^9 + O(q^{10}).\e$$
 
 From the program ALGEBRA we find the string functions in the affine algebra representation,
 $$q^{d_0} c^{\Lambda_1}_{\Lambda_1}(\tau)=
1 + 4 q + 16 q^2 + 50 q^3 + 143 q^4 + 368 q^5 + 892 q^6 + 2035 q^7 + 
 4448 q^8+O(q^9),\e$$
 and 
 $$q^{d_1} c^{\Lambda_1}_{\Lambda_1+\alpha_1}(\tau)=
 1 + 6 q + 22 q^2 + 70 q^3 + 193 q^4 + 493 q^5 + 1170 q^6 + 2642 q^7+O(q^8),\e$$
 where $d_\mu$ are some dimensions which folkows from eq. (4).
 Multiplying these string functions by $\eta(\tau)^2$, to get the characters, we find exactly
 the expressions calculated from the GRR, eqs. (35,36).
 
 We studied many other string functions for this algebra, confirming our GRR formula for the algebra $A_2$ at level $3$.
 
 One may check in this fashion other algebras. We verified our GRR formula, eq. (18),
 for many algebras at many levels. These include the non--simply laced algebras,
 $B_2$, $B_3$, $C_3$, $G_2$, $F_4$ at level two,  and the simply laced algebras
 $A_2$ at levels $3,4$, $A_3$, $D_4$ and $E_6$ at level $3$. We checked in these algebras
 many of the string functions and, indeed, the GRR expressions agree exactly with the string 
 functions, as computed by ALGEBRA. 
 
 We described  here a Rogers Ramanujan type expressions for the characters of 
 parafermionic cosets of the type $H_l/U(1)^r$
 for all the algebras $H$ at level $l$, where $r$ is the rank. The characters are $\chi^\Lambda_\lambda$, 
 where $\Lambda$ is an integrable highest weight, and $\lambda$ is a weight. Our conjecture
 holds for all $\Lambda$ such that $\Lambda=f\Lambda_g$, where $f$ is an integer,
 and $\Lambda_g\cdot\theta=1$ where $\theta$ is the highest root, i.e., $\Lambda_g$ has mark
 one, and $\alpha_g^2=2$. Our conjecture holds also for short roots, which have dual mark one,
 $\alpha_g^2<2$,
 provided $f=1$. Our GRR expression holds for all the  weights  $\lambda$. 
 We verified this conjecture using Freudenthal Kac formula, which we computerised
 in the fortran program AGEBRA. This program gives the exact characters, up to some grade.
 
 It is an interesting question to generalise  our GRR expressions to the weights $\Lambda$
 other than the ones above. Another interesting question is how to connect these expressions
 with solvable lattice models of the RSOS type (rigid solid on solid). This will, also, furnish a physical
 proof for these identities. 
 
 The parafermionic  characters are related to the string functions. The string functions are central
 in the theory of affine algebras. Thus, our expressions, 
 which give the first closed formula for these new string functions,
 are also very important mathematically.

\ack
I would like to thank A. Abouelsaood for his collaboration in writing the fortran program ALGEBRA.
\refout
\bye